\documentclass[global,twocolumn]{svjour}
\usepackage{graphicx,float}
\begin{document}
\title{Population trapping and inversion in ultracold Fermi gases by
  excitation of the optical lattice - Non-equilibrium Floquet-Keldysh description}
\author{Regine Frank%\inst{1} %\and Second author\inst{2}% etc
}                     % Do not remove
\institute{Insitute for Solid State Physics, Karlsruhe Institute of Technology
  (KIT), Wolfgang-Gaede-Strasse 1, 76131 Karlsruhe \\
{\it present address:} Institute for Theoretical Physics, Eberhard-Karls Universit\"at
           T\"ubingen, Auf der Morgenstelle 14, 72076 T\"ubingen, Germany}
%\institute{ \and }
\date{Received: 17.12.2012 / Revised version: date}
\maketitle
\begin{abstract}
A gas of ultracold interacting quantum degenerate Fermions is considered in a 
three dimensional optical lattice which is externally modulated in the
frequency and the amplitude. 
This theoretical study utilizes the Keldysh formalism to account for  
the system being out of thermodynamical equilibrium. A dynamical  
mean field theory, extended to non-equilibrium, is presented to calculate 
characteristic quantities such as the local density of states and the 
non-equilibrium distribution function. A dynamics Franz-Keldysh splitting is
found which accounts for the non-equilibrium modification of the underlying
bandstructure.
% and it leads to a description of a phonon driven atom laser below the threshold. 
The found characteristic Floquet-fan like bandstructure
accounts for the quantized nature of the effect over all frequency space.

\end{abstract}

\section{Introduction}
\label{introduction}
Non-equilibrium physics is often counter-intuitive at first instance
and additionally has often been neglected due to the sophisticated methods
needed in order to theoretically describe such phenomena. This is even more 
unfortunate, given that non-equilibrium surrounds us every day in real life,
e.g. in hydrodynamic flows etc.
Let's start with some simple thoughts to explain the formalism: If
we consider as 'toy-model' e.g. a simple two body problem, it is well known
that one can transform the whole system in a center-of-mass problem, and a
problem which refers to the relative coordinates with respect to that
center-of-mass. So far everything is straightforward. If one considers
equilibrium physics in the interaction of the two bodies, all processes can be
described in terms of the relative coordinates. The center of mass is resting
in space and time, or it is moving uniformly. When we consider
non-equi\-librium processes, the situation changes significantly. 
The
center-of-mass may exhibit a distinctly different behavior in driven,
i.e. non-equilibrium, systems.
Furthermore, the number of equations or sets of equations required to
adequately describe a given system, is usually increased. Partially and among
other reasons this is
also due to account for the dynamics of the center-of-mass.
Because of the center-of-mass dynamics, there may be a significant amount of
energy   'stored' or contained in these dynamics. 
That amount of stored
energy consequently does not affect the two body body system directly, to be
more precise an imaginary observer on either of the two bodies will not 'see' 
the results due to the center-of-mass dynamics as he is affected by the
dynamics of relative coordinates. In this imaginary setup, the outer world is
however also sensitive to the center-of-mass dynamics, which may change the
system behavior significantly, as compared to systems without such center-of-mass dynamics.\\

When we consider ultracold gases, as investigated in various exciting experiments
\cite{Ketterle,Bloch,LisaGroup01,LisaGroup02,LisaGroup03,HamburgBarcelonaGroup01,HamburgBarcelonaGroup02,HamburgBarcelonaGroup03} 
in optical lattices 
in non-equilibrium, we
take exactly the point of view into account which is described above: The
behavior of the atoms in their own frame of reference is not affected by
non-equilibrium, they obey the physics with respect to their equilibrium groundstate.
Alas the band structure, the dispersion relation which can be
measured by the observer from outside is significantly changed. The system all
together in non-equilibrium takes a new groundstate as reference, and the
equilibrium ground state becomes irrelevant. In consequence, that fact leads
e.g. to a significant aberrance of the ground state from the Fermi edge for
ultracold Fermions in optical lattices even as a state of half-filling is
considered. Furthermore, it is clear that the equilibrium state is not to be reached
'smoothly' from an excited state by decreasing the energy, but that process
represents, in fact, a phase transition.\\
Within this article an ultracold fermionic manyparticle quantum system on a
lattice is studied. The lattice is modulated by vibrations and the
dynamical response is calculated by raise of the modulations' amplitude for
various frequencies. This systems are the most ideal setups to study Hubbard
physics in non-equilibrium setups.  

\section{Model and Theory}
\label{model_and_theory}

The physical system under consideration is comprised of an 
optical lattice in three dimensions (3D) in which an ultracold 
atomic gas has been loaded. The gas consists of fermionic atoms
with a strong repulsive on-site interaction. Due to the interaction a
strongly correlated system is formed. The 3D optical lattice, 
formed by pairs of counter-propagating laser beams, is externally 
tuned such that the potential strength of the formed lattice 
changes in a highly controllable way. Consequently, the intrinsic 
properties of the lattice, such as hopping amplitude or on-site
energy, change accordingly. The modulation of the lattice shall
be oscillatory in time but otherwise constant. This means the 
potential depth experienced by an atom at a given site oscillates
around fixed value.

The considered system of ultracold fermions placed in an optical lattice
with  temporal modulations can be described by the following Hamiltonian
\begin{eqnarray}
\label{Eq:Hamilton}
H (\tau) 
&=&
\sum_{i,\sigma} \left(\epsilon_0+\widetilde{E}(\tau)  \right)
c^{\dagger}_{i,\sigma}c_{i,\sigma} \\ \nonumber
\,\,\,\,-\!\!\!\!\sum_{\langle i,j \rangle, \sigma}\!\!\!\!&&\!\!\!\!\left(t+\widetilde{T}(\tau)\right)
c^{\dagger}_{i,\sigma}c_{j,\sigma}  +
\,\frac{U}{2} \,\, \sum_{i,\sigma}
c^{\dagger}_{i,\sigma}c_{i,\sigma} c^{\dagger}_{i,-\sigma}c_{i,-\sigma}
\end{eqnarray}
where the first term on the right hand side (rhs) represents
the on-site energy of the atoms, the second term on the rhs 
the kinetic or hopping contribution between nearest neighbor 
lattice sites and the last term on rhs represents the repulsive 
interaction with strength $U$ between two atoms located at the 
same lattice site. The letter $\tau$ marks  the temporal 
dependence on time, $\epsilon_0$ is the equilibrium or 
static on-site energy, whereas $\widetilde{E}(\tau)$ is the time 
dependent contribution or modulation of this on-site energy. 
The hopping in this time dependent model is modified likewise, 
$t$ is the regular hopping amplitude of an atom between two 
adjacent lattice sites and  $\widetilde{T}(\tau)$ is its time
dependent modification. If two atoms reside at the very same 
lattice site they encounter a repulsive interaction of strength 
$U$, this interaction strength is however not to be modified by 
temporal changes due to its overwhelming strength compared to 
possible temporal changes. The operators $c^{\dagger}_{i,\sigma}$  
and  $c_{i,\sigma}$ create and annihilate an atom at lattice site 
$i$ with spin $\sigma$ in Eq. (\ref{Eq:Hamilton}). 
Furthermore, the symbol $\langle i,j \rangle$ 
refers to a sum over nearest neighbors only. 
 
Throughout this paper, the time dependent modulations of the 
lattice, and therefore also of the on-site energies and hopping  
amplitudes, are assumed to be periodic in time. Consequently,  
the time dependent contributions in Eq. (\ref{Eq:Hamilton}) are  
assumed to be of the form 
\begin{eqnarray} 
\label{Eq:Parameter} 
\widetilde{E}(\tau)&=& E \cos\left(\Omega \tau \right)  \\ 
\widetilde{T}(\tau)&=& T \cos\left(\Omega \tau \right) 
\end{eqnarray} 
where  $\Omega$ represents the frequency which is used to  modulate 
the lattice. A careful choice of parameters $E$ and $T$  
guarantee that no sign change as a function of time $\tau$ will  
occur in the Hamiltonian $H(\tau)$, Eq. (\ref{Eq:Hamilton}).

At this point it is worth noting two important points concerning 
the theory. 
First, due to the external driving of the optical  
lattice, the system can not reside in a state of thermodynamical  
equilibrium and therefore requires special care, intrinsic  to  
the treatment of nonequilibrium systems, as e.g. proposed by  
Schwinger and Keldysh \cite{Keldysh}. 
In a nonequilibrium system the current state and also  
physical quantities, such as e.g. the density of states, depend  
always on two time arguments, for instance a starting time and  
the elapsed time, or equally a relative time and a 
center-of-mass time. In contrast, in equilibrium theory, there is  
usually only one time coordinate needed, the relative time, since  
the center of mass time can not change any physics. However,  
in the considered nonequilibrium system, physical quantities  
depend on two time arguments. An appropriate formalism to consistently describe
such systems using e.g. diagrammatic descriptions, has been developed first by
Schwinger and then by Keldysh  \cite{Keldysh}, 
by introducing a Green's function according to
\begin{eqnarray}
G (\tau_1,\tau_2) = 
\left(
\begin{array}{lcl}
G^{++}(\tau_1,\tau_2)   & \qquad &  G^{+-}(\tau_1,\tau_2)   \\
G^{-+}(\tau_1,\tau_2)   & \qquad &  G^{--}(\tau_1,\tau_2)
\end{array}
\right)
\end{eqnarray}
where the superscripts depend on which branch 
of the Schwinger-Keldysh-Schwinger contour the time 
arguments reside. To give an explicit example, $G^{++}(\tau_1,\tau_2)$ refers to a
situation, where $\tau_1$ and $\tau_2$ are both the upper ($+$) countour
path.
By using a rotation $R$, given by 
\begin{eqnarray}
R = \frac{1}{\sqrt{2}}
\left(
\begin{array}{lr}
1  &   -1    \\
1  &   1
\end{array}
\right),
\end{eqnarray}
in this Keldysh space, the Keldysh Schwinger Green's
function may also be written in the form
\begin{eqnarray}
G (\tau_1,\tau_2) = 
\left(
\begin{array}{lcl}
0                     & \qquad &   G^{adv}(\tau_1,\tau_2)   \\
G^{ret}(\tau_1,\tau_2)   & \qquad &   G^{keld}(\tau_1,\tau_2)
\end{array}
\right)
\end{eqnarray}
where $G^{adv}$ and $G^{ret}$ are the well known advanced and retarded
components of the Green's function, and the new  $G^{keld}(\tau_1,\tau_2)$ represents
the so-called Keldysh component of the Green's function. As in equilibrium
physics, the advanced and retarded parts are connected to the actual state of
the system, i.e. its spectral weight and the density of states, whereas the Keldysh component
additionally includes   information on the nonequilibrium distribution. Especially the
latter point will utilized later on to find the according non-equilibrium
distribution function.
The second point to note here is the strictly  
periodic nature of the modulation, which allows for the use of  
the so-called Floquet theory  \cite{Floquet}. In order to  
take full advantage of these conditions, it is appropriate  
to change to Fourier space and using a momentum and frequency  
description.  
As detailed above, the description of  nonequilibrium systems
requires the use of two independent time arguments
and consequently requires a two-time Fourier  
transform towards the corresponding  two-frequency expression.  
The relative frequency $\omega$ is the very same as known from  
equilibrium theory and the center of mass frequency corresponds  
to the modulation frequency of the optical lattice $\Omega$. 
In particular, the employed Fourier transform reads
\begin{eqnarray}
G^{\alpha\beta}_{mn}(k,\omega,\Omega)
&\!=\!&
\int_{-\infty}^{+\infty}\!\!\!
{\rm d} \tau_{rel}
\frac{1}{\mathcal{T}} 
\int_{ -\mathcal{T} /2}^{+\mathcal{T}/2}\!\!\!
{\rm d}\tau_{cm} 
e^{i(\omega−\frac{m+n}{2} \Omega_L)\tau_{rel} } \nonumber\\
&& \times 
e^{   i(m−n)\Omega_L \tau_{cm}  }
G^{\alpha\beta}  (k, \tau_{rel} , \tau_{cm}  )
\label{Eq:Fourier}
\end{eqnarray}
where the Greek superscripts denote the Keldysh indices, i.e.  
$\alpha$ and $\beta$ may assume the values $+$ or $-$ depending  
on which branch of the Keldysh contour the time arguments reside. 
The subscripts $m$ and $n$ label the Floquet index, i.e. they  
account for the number of absorbed or emitted lattice quanta  
$\hbar\Omega$. The quantity $\mathcal T$ in Eq. (\ref{Eq:Fourier})  
is the time period $ \mathcal{T} = \frac{2 \pi}{\Omega}$, the system 
time is shifted to a center of motion time  
$\tau_{cm}= \frac{\tau_1+\tau_2}{2}$ and a relative time coordinate 
$\tau_{rel} = \tau_1 - \tau_2$. 
The resulting Keldysh Schwinger Floquet Green's function is therefore a matrix
Green's function of dimension $2\times 2$ in Keldysh space and of dimensions
$(2n+1)\times (2m+1)$ in Floquet space. In numerical evaluations, a size of $21\times
21$ has been used in Floquet space, corresponding to a consideration of
emission and absorption of up to 10 lattice quanta at the same time (10 phonon
processes). 
Combinations of these techniques Schwinger-Keldysh technique with the Floquet
formalism have been considered for some time see e.g. \cite{Jauho,Aoki,Sau}.

The Hamiltonian, Eq. (\ref{Eq:Hamilton}), including the  
finite on-site interactions, is solved at zero temperature  
and at half filling of the lattice sites by extending   
an equilibrium dynamical mean field theory (DMFT), see for  
instance reference \cite{Georges}, towards accounting for  
this  nonequilibrium situation by including the Floquet-Keldysh  
Green´s function described in Eq. (\ref{Eq:Fourier}). 
This  DMFT maps the non-equilibrium interacting lattice system onto  
a local impurity system, still out of thermal equilibrium,  
by assuming a local on-site selfenergy. 
This is in some way comparable to the light-matter interaction considered in
ref. \cite{Aoki}.
The DMFT provides a 
solution for the derived local impurity problem.  
The iterated perturbation theory (IPT) \cite{Zhang}  
has been extended to a nonequilibrium description. 
This method as a diagrammatic  
impurity solver is reasonable fast also for  
nonequilibrium systems, because there exist analogues \cite{Keldysh}  
to the Feynman rules for evaluating equilibrium diagrams.  
Especially at half filling, the IPT proves to  be a reliable  
solver for the DMFT in equilibrium systems \cite{Zhang}. 
 
At the end one is left with the task of selfconsistently  
evaluating a matrix equation which is of dimensions 2x2 in Keldysh  
space and of dimension  $(2n+1)\times (2n+1)$ in Floquet space.  
In this paper  $n=10$ Floquet bands have been used in the  
numerical evaluations, i.e. $n$ ranges from $-10$ to $+10$.  
Practically, this number of Floquet bands corresponds to the  
number of emitted or absorbed lattice quanta $\hbar \Omega$. 
The numerical solution of this nonequilibrium DMFT approach, leads  
then to the full  non-equilibrium Floquet-Keldysh-Green’s  
function Eq. (\ref{Eq:Fourier}), i.e.  to the full knowledge 
of the three distinct components of the
local Green's function $G^{ret}(\omega,\Omega)$, $G^{radv}(\omega,\Omega)$ and
as well as $G^{keld}(\omega,\Omega)$ for all atomic energies $\hbar\omega$ and all
lattice vibrations $\hbar\Omega$.
Therefore, also to physical quantities such as the  
local density of states or the nonequilibrium distribution  
function of the ultracold fermionic gas in the modulated optical  
lattice. In particular the local density of states $N(\omega,\Omega)$ is given by  
the expression 
\begin{eqnarray}
\label{Eq:DOS}
N(\omega,\Omega) 
&=& 
-\frac{1}{\pi}{\rm Im\,}G^{ret}(\omega,\Omega) \\
&=&  
-\frac{1}{\pi}\int {\rm d}^3k  \sum_{m,n,\sigma} {\rm Im\,} G^{ret}_{mn;\sigma}(k, \omega,\Omega)\nonumber
\end{eqnarray}
since all emission and absorption processes have to be included, 
there appears a sum over all involved Floquet bands ($\sum_{mn}\ldots$) in Eq. (\ref{Eq:DOS}).

\begin{figure}[t]
\resizebox{0.51\textwidth}{!}{%
\includegraphics[clip]{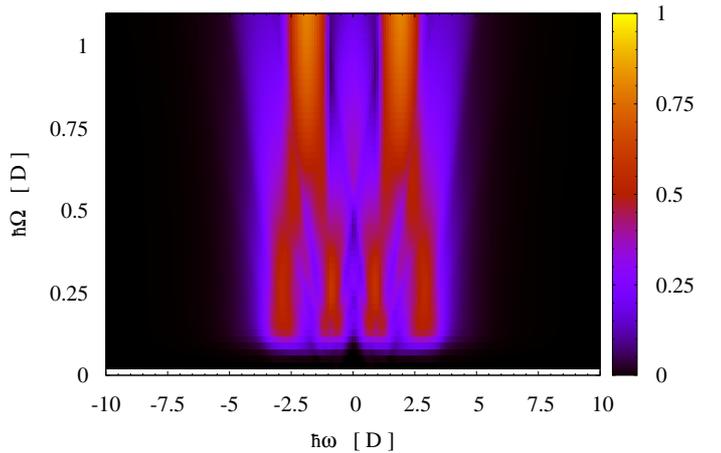}%
}
\caption{Imaginary part of the retarded component of the full and local Green's
  function. This relates to the density of states 
  (LDOS) $N(\omega, \Omega)$, as given in Eq. (\ref{Eq:DOS}).  
  The abscissa represents the atomic 
  energy $\hbar\omega$, whereas the ordinate $\hbar\Omega$ labels the energy of the
  lattice modulation, i.e. its frequency. Here and in the following the unit
  of energy is given by the half bandwidth $D$ of the equilibrium system.
  The interaction strength of two atoms with opposite spin at the same lattice
  site is $U/D=4$ and the modulation hopping amplitude is here set to be
  $T/D=0.25$.  For a detailed discussion see text. }
\label{Fig:01}     
\end{figure}

\section{Results and Discussion}
\label{results_and_discussion}

\begin{figure}[t!]
\resizebox{0.51\textwidth}{!}{%
\includegraphics[clip]{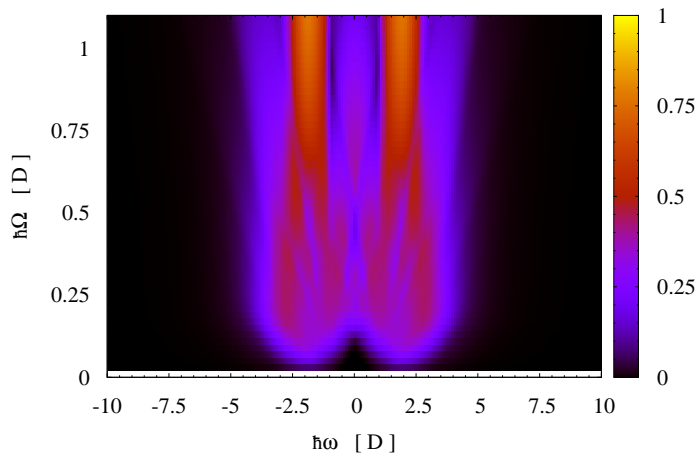}%
}
\caption{ The same parameters as in Fig. \ref{Fig:01} but with the modulation induced hopping
  increased to a value of  $T/D=1.00$. 
  For a detailed discussion see text below.}
\label{Fig:02}     
\end{figure}

In this section results of the numerical evaluation of the  
nonequilibrium DMFT are presented for a fixed set of on-site  
energy and on-site interaction strength and for a range of  
modulation strengths $T$ of the optical lattice. The on-site energy 
variation is here set to be $E/D=1$. 
Here throughout this paper the unit of energy is given by the half 
bandwidth $D$ of the equilibrium system.
In particular,  
the system is at half filling, the on-site  
interaction is set to be $U/D=4$, i.e. it is a strongly 
interacting system. The equilibrium ground state of this  
system is therefore an insulating state, i.e. there is no  
spectral weight at the Fermi energy $\hbar\omega=0$. The two bands,  
i.e. the lower and the upper Hubbard band, are split by $U/2$  
due to the strong correlations. In other words, a gap of width $U/2$  
opens up in the spectral weight, consequently conductivity is  
suppressed and a Mott insulating state is established. The  
energy gap occurs symmetrically in the presented plots, because  
in the calculations the filling of the optical lattice is  
assumed to be $0.5$, i.e. each lattice site is occupied by one atom.  
In the presence of periodic lattice strength variations, i.e.  
in the non-equilibrium regime, this situation is altered.  
With increasing strength of the lattice  modulations and depending  
on the modulation frequency $\Omega$ this changes in some respects, 
as detailed below.

\begin{figure}[t!]
\resizebox{0.5\textwidth}{!}{%
\includegraphics[clip]{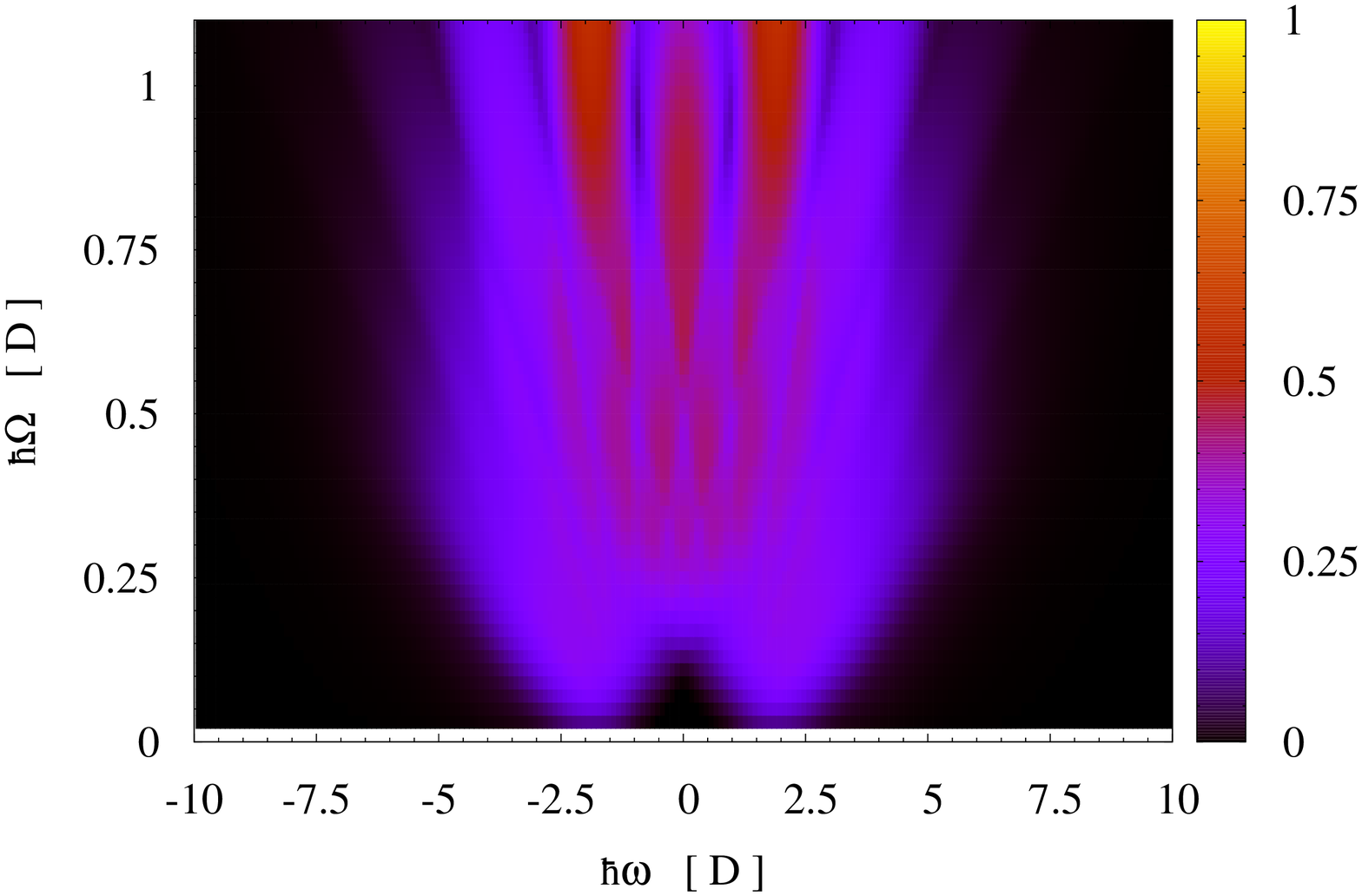}%
}
\caption{The same parameters as in Fig. \ref{Fig:01} but with the modulation induced hopping
  increased to a value of  $T/D=3.00$.  For a discussion see text.}
\label{Fig:03}     
\end{figure}

\begin{figure}[t]
\resizebox{0.5\textwidth}{!}{%
\includegraphics[clip]{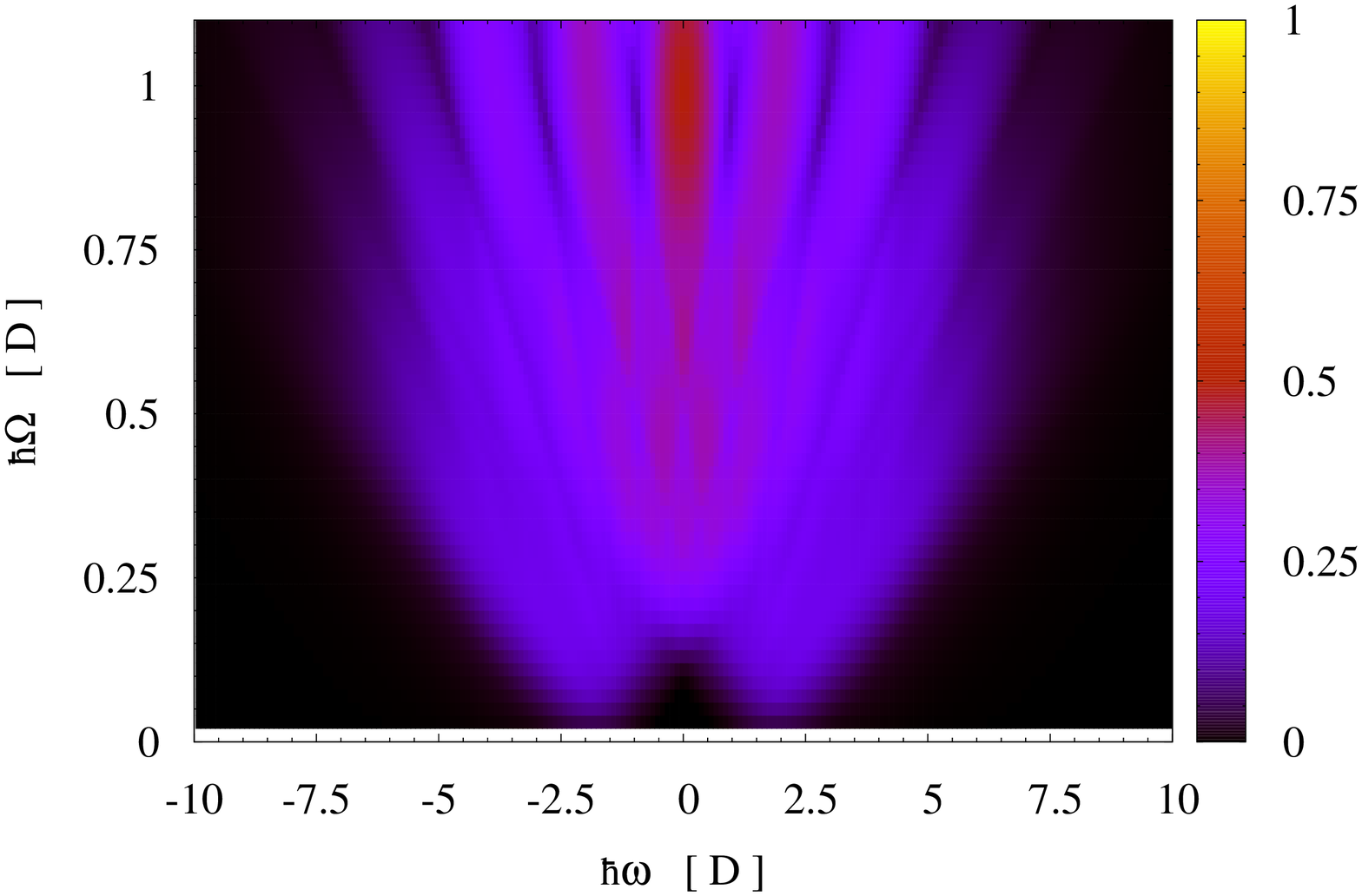}%
}
\caption{The same parameters as in Fig. \ref{Fig:01} but with the 
  modulation induced hopping
  increased to a value of  $T/D=5.00$.  For a detailed discussion see text below.}
\label{Fig:04}     
\end{figure}

\begin{figure}[t]
\resizebox{0.5\textwidth}{!}{%
\includegraphics[clip]{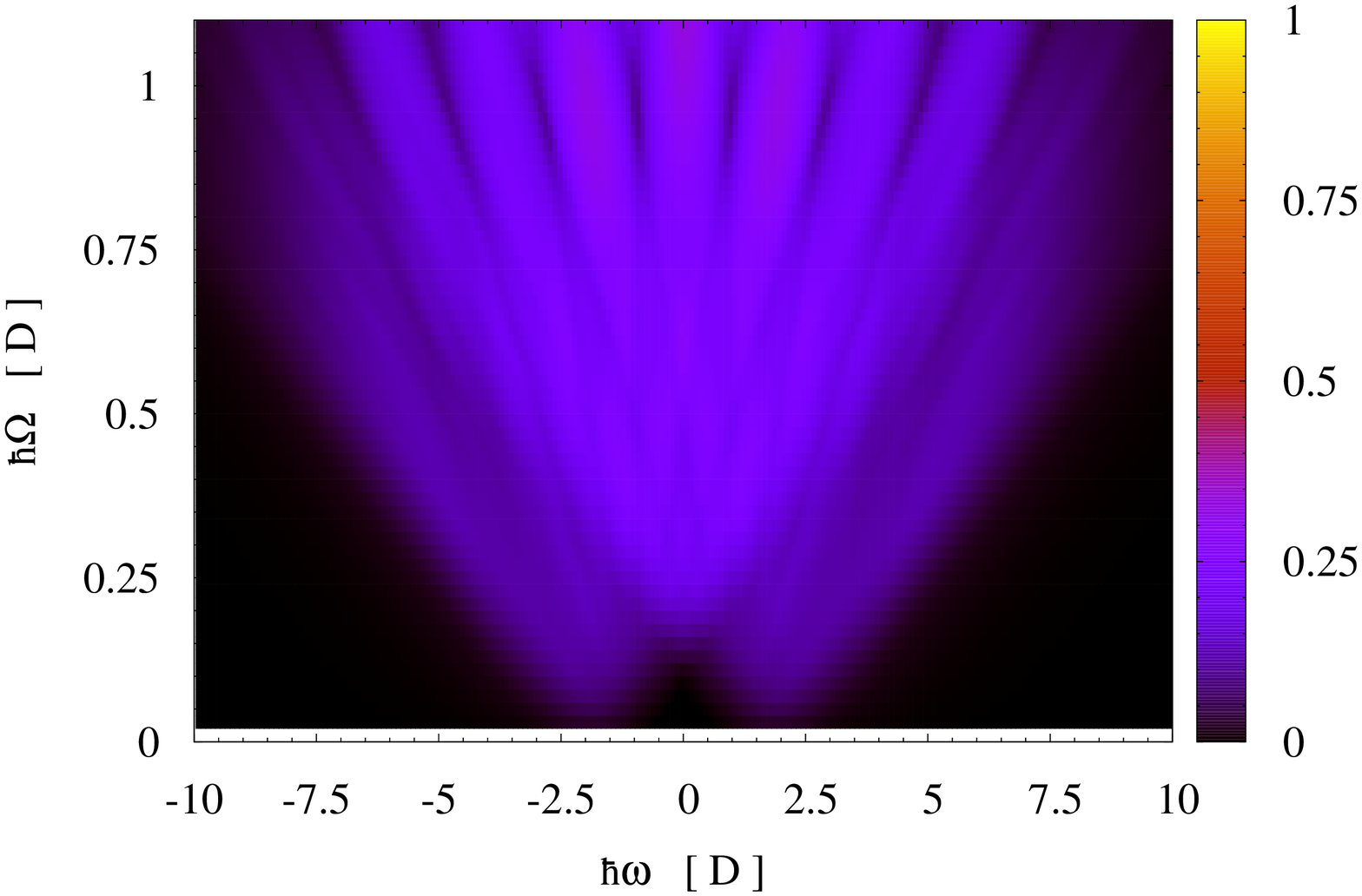}%
}
\caption{The same parameters as in Fig. \ref{Fig:01} but with the 
  modulation induced hopping
  increased to a value of  $T/D=8.00$. 
  For a detailed discussion see text below.
}
\label{Fig:05}     
\end{figure}

\begin{figure}[t!]
\resizebox{0.5\textwidth}{!}{%
\includegraphics[clip]{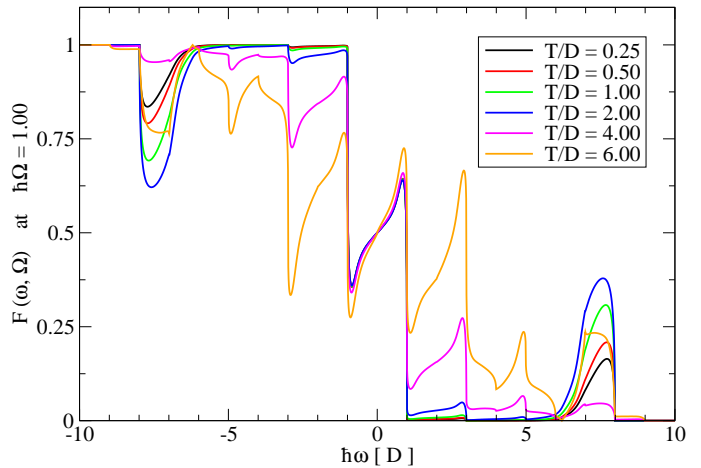}%
}
\caption{Numerically computed non-equilibrium distribution function $F(\omega,
  \Omega)$ of the
dynamical AC-Wannier-Stark effect of fermionic atoms as a 
function of atomic energy $\hbar\omega$ for the modulation
frequency of  $\hbar\Omega/D=1.0$. Clearly visible are the absorption and emission
features of lattice modulation quanta, a spectral hole burning effect. The
absorption features for $\hbar\omega/D<0$ display are sharp lower edge, since there
is a specific maximum energy for each absorption process. The different
absorption features correspond to the different number of involved lattice
quanta. Processes involving simultaneous absorption of up to four lattice
quanta are found.}
\label{Fig:07}     
\end{figure}

\begin{figure}[t!]
\resizebox{0.5\textwidth}{!}{
\includegraphics[clip]{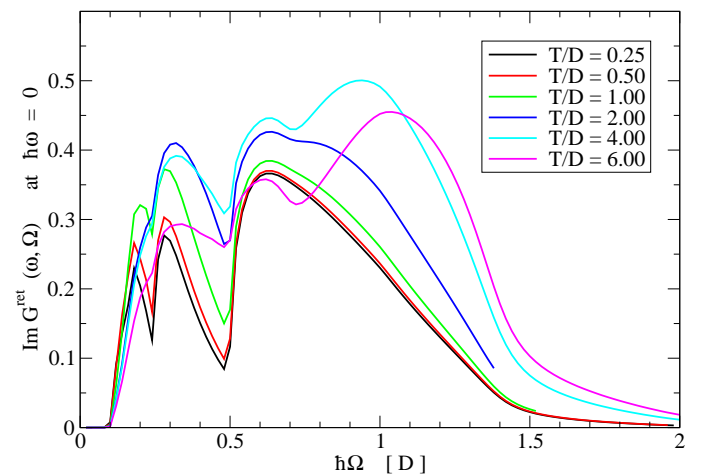}%
}
\caption{Displayed is the computed spectral weight 
${\rm Im\, G^{ret}(\omega, \Omega)}$, c.f. Eq.
(\ref{Eq:DOS}), along the Fermi edge at $\hbar\omega/D=0$ as a function of lattice
modulation frequency $\hbar\Omega$. Different curves correspond to different
modulation strengths as indicated. The density of states at the Fermi edge is
also an indicator of the conductivity of system. Strong non-monotonic
variations of the spectral weight are observed depending on both, modulation
strength and frequency. As universal features, there appears a minimum at ca.
$\hbar\Omega/D=0.5$ followed by a maximum at ca. $\hbar\Omega/D=0.65$. Interestingly, for
small and large modulation frequencies the spectral weight returns to zero,
its equilibrium value.}
\label{Fig:06}     
\end{figure}

In Figs. \ref{Fig:01} to \ref{Fig:06} the local density of states
is displayed for a series of increasing amplitudes of the periodic 
optical lattice modulation. Each plot itself features the 
atomic energy $\hbar\omega$ 
along the abscissa and the energy $\hbar\Omega$, i.e. frequency, of the lattice 
modulations along the ordinate. So, for a specified lattice vibration 
frequency, one chooses that specific value at the y-axes and reads out 
the density of states as a function of atomic energy $\hbar\omega$ along the 
x-axes. 

For small external modulation frequencies the original bands are 
found to split into a number of Floquet-sidebands which are separated to 
each other proportional to the external modulation frequency, therefore 
the almost equally smeared-out appearance for small vibrational frequencies, 
see e.g. Fig. \ref{Fig:01}. For this example of a small amplitude 
modulation $T/D=0.25$, another interesting effect is to be observed, 
starting at about modulation energies of $\hbar\Omega/D=0.12$.

The interaction-based splitting of the bandstructure of upper and 
lower Hubbard band in Fig. \ref{Fig:01} is modified in the way that the two Hubbard 
bands themselves are again split into two, such that an effective 
four-band system is formed. This additional splitting is however not caused
by atom-atom interactions but by the dynamical modulation of the lattice 
potential. This newly established state  still
maintains, at least  approximately, the insulating character of the underlying 
equilibrium solution, i.e. no spectral weight and atomic occupation 
at the Fermi edge, i.e. at $\hbar\omega=0$. 
These four formed quasi-bands are separated by regions 
of low spectral weight, so to speak by quasi-gaps which opened 
solely due to externally induced lattice vibrations in this ultracold 
Fermi-gas system. In a modulation range between $\hbar\Omega/D=0.4$ 
and $\hbar\Omega/D=0.6$ where the 
fast modulations of the lattice only weakly affect the local 
density of states of the degenerate atomic gas.

The decreasing effect of the dynamic lattice vibrations on 
the density of states with increasing frequencies $\Omega$ is attributed to  
the fact that at some point the atoms in the optical lattice cannot follow the 
quickly changing lattice potential anymore. Therefore, a local density  
of states is formed which is in its main features similar to the equilibrium  
density of states, compare to   Fig.  \ref{Fig:01} and vibration frequencies 
of ca. $\hbar\Omega/D \ge 1$. In this regime, only one excited Floquet band of both  
upper and lower Hubbard band is still visible, with minor deviations inside  
the equilibrium gap, i.e. between ca. $-1<\hbar\omega/D<+1$. 
Each Hubbard band forms a Floquet sideband  of emission of one vibrational 
quant $\hbar\Omega$ to its left and one  
Floquet sideband   of absorption of one vibrational quant $\hbar\Omega$  to its  
right hand side.

In Figs. \ref{Fig:02}-\ref{Fig:05}, we present numerical results for the LDOS,
Eq. (\ref{Eq:DOS}), for numerically ascending modulation induced hopping
strengths of $T/D = 1,3,5,8$. The spectral weight for intermediate strengths $T/D$,
Fig. \ref{Fig:02} and Fig. \ref{Fig:03}, the range of small modulation
energies $\hbar\Omega \le 0.25$, displays the change from modified equilibrium
DOS towards a split up two band system, split into a quite large number of
Floquet side bands. The synthetic, generated quasi four-band system is still
found for approximately  $\hbar\Omega/D \le 0.65$. For even larger modulation
energies in  Fig. \ref{Fig:02} and Fig. \ref{Fig:03} the 
spectral weight assembles again the
equilibrium two Hubbard bands and to some extent an 
additional spectral weight in the gap region  
between $1 \le \hbar\omega/D \le 1$. For quite large hopping modulations of 
$T/D =  5$ and $ T/D = 8$ presented in Fig. \ref{Fig:04} and Fig. \ref{Fig:05}, 
the just described characteristics are absent and replaced by a Floquet-fan like
structure. This found structure represents the Floquet sidebands of absorption
and emission of vibration quanta of the lattice. Interesting to note about 
Fig. \ref{Fig:04} and Fig. \ref{Fig:05} is that the structure of the LDOS for
large  $\hbar\Omega$ displays a reversed structure as compared to the
equilibrium system. Precisely, in the former gap region  
$1 \le \hbar\omega/D \le 1$, there is significant spectral weight, whereas in
the place of the former Hubbard bands the spectral weight is suppressed.

In general, this formed structure in the spectral function resembles a Wannier-Stark ladder, 
which is also seen 
in the energy distribution of the fermionic atoms in the lattice.
The non-equilibrium distribution function $F(\omega,\Omega)$ of the ultra-cold and strongly 
interacting Fermi gas is obtained with the
help of the Keldysh component\\  $G^{Keld}(\omega, \Omega)$ of the full Keldysh-Floquet-Green's 
function by means of the following general relation
\begin{eqnarray}
F(\omega,\Omega) = \frac{1}{2}\left(1+\frac{1}{2i} 
\frac  {\sum_{mn}G^{Keld}_{mn}(\omega,\Omega)}
       {\sum_{mn}{\rm Im\,} G^{ret}_{mn}(\omega,\Omega)}\right).
\end{eqnarray}

A numerical evaluation of this distribution function, characterizing the 
atoms inside the periodically modulated lattice, is depicted in 
Fig. \ref{Fig:07}. The distribution function $F(\omega, \Omega)$
is depicted for various modulations strengths $T$ as a function of atomic
energy $\hbar\omega$ for one specific value of the modulation frequency 
$\hbar\Omega/D=1$. Clearly visible are the absorption and emission
features of lattice modulation quanta, a spectral hole burning effect. The
absorption features for $\hbar\omega/D<0$ display are sharp lower edge, 
since there is a specific maximum energy for each absorption process.

The different
absorption features correspond to the different number of involved lattice
quanta. For instance, atoms that absorb energy at $-\hbar\omega$ are lifted in energy 
to a state with energy $+\hbar\omega$ by absorbing energy from the lattice 
oscillation with $2\hbar\omega=n\hbar\Omega$ where n characterizes the number of
absorbed lattice quanta. Processes involving simultaneous absorption of up to 
four lattice quanta are found. The process of transferring energy to the
lattice by emitting lattice quanta is analog. In Fig. \ref{Fig:07} the 
at first glance overproportional dip (peak) in the distribution function 
at atomic energies $\hbar\omega/D \sim -8$ ($\hbar\omega/D\sim +8$) 
is best understood 
when bearing in mind that the physical quantity involved here is the atomic 
occupation number, which is the product of distribution and spectral function. 
Since this effect in the distribution function occurs in a region where 
the spectral function, see for instance Fig \ref{Fig:04}, is weak, the 
occupation number is only slightly modified by this process.

Returning now to the discussion of the density of states.
With increasing modulation strength $T$ of the optical lattice  
holding the ultracold atomic gas, the distortion of the equilibrium  
density of states becomes more and more severe. By comparing  
Figs. \ref{Fig:02}-\ref{Fig:04} one observes that the equilibrium  
density of states is almost completely changed. This is due to  
the formation of an increasing number of  Floquet sidebands. 
These absorption and emission bands  may also encounter intersections 
and fill even the Hubbard gap. This transferring   
of spectral weight towards atomic energies $-1<\hbar\omega/D<1$ leads to  
the breakdown of the Mott insulating state and consequently the  
system selfconsistently passes over into a conducting state.  

With an even more increased modulation amplitude $T$, see e.g.  
Fig. \ref{Fig:05}, the system undergoes a transition into a state  
in which for rather large modulation frequencies ca. $\hbar\Omega/D >0.8$  
the system almost resembles a seven-band system with the zeroth band  
right around the Fermi-energy $\hbar\omega/D=0$, even though the system  
is severely driven out of equilibrium. The number and position of the  
Floquet sidebands is also highly sensitive to driving frequency  
and may change drastically, depending on the specific value of $\hbar\Omega/D$,  
cf. Fig. \ref{Fig:05}.  
Interestingly, this behavior opens a way to switch the system in 
a controlled way from an insulating multi-band system to a conducting 
multi-band system, by changing the modulation amplitudes from low to high at 
a fixed external modulation frequency.

The density of states at the Fermi energy is always of special interest, since
it is a reliable indicator  for the conductivity of the system in question. 
In Fig. \ref{Fig:06} the local density of states at the Fermi edge is depicted
for a variety of modulation strengths. Different curves correspond to different
modulation strengths as indicated. These curves therefore correspond to
vertical cuts along $\hbar\omega/D=0$ in Figs. \ref{Fig:01}-\ref{Fig:05} and are 
presented in their own graph simply for the reasons of clarification and clearness.
Strong non-monotonic
variations of the spectral weight are observed depending on both, the modulation
strength and frequency. As universal features, there appears a minimum at ca.
$\hbar\Omega/D=0.5$ followed by a maximum at ca. $\hbar\Omega/D=0.65$. For
small and large modulation frequencies the spectral weight returns to zero,
which is its equilibrium value, since the corresponding 
ground state is a Mott insulator. In general, two-phonon processes which are 
bridging the gap at modulation energies of $\hbar\Omega=1$ yield a strong
conductivity in this driven system.

\section{Summary and Conclusion}
\label{conclusion}
	
In conclusion, an ultracold gas of strongly interacting fermionic atoms placed in a  
three-dimensional optical lattice has been considered. The potential strength of the  
optical lattice has been periodically modulated in time. Due to these external  
modulations the gas is driven out of  thermodynamical equilibrium. A theoretical  
analysis of the nonequilibrium state was performed by employing the 
Schwinger-Keldysh technique in order to account for far off equilibrium
situation and in combination with  a Floquet analysis, best suited to take 
advantage of the periodic nature of the driving. The system has been solved by
a DMFT method, developed to include the nonequilibrium  
Floquet-Keldysh formalism and solved by means of the IPT. The IPT impurity 
solver has  also been extended to account for nonequilibrium  system.

The developed technique was applied to calculate the  local density of states of  
the driven system of an ultracold Fermi-gas in a modulated three-dimensional optical  
lattice. The influence of a systematic change of the externally 
controlled amplitude of the lattice modulation was studied.
The numerically determined LDOS displays severe changes depending 
on both the driving frequencies and the modulation amplitude. 
The strongly correlated system can be driven out of its 
insulating state into a conducting multi-band dynamical Wannier-Stark state with quite interesting 
effects in the intermediate regime.
The observed inversion as compared to the equilibrium state, 
i.e.  the occupied states above the lower Hubbard band
and consequently the lowered population below, is open to experimental
verification. In particular the part of the gap region 
$0 \le \hbar\omega/D \le 1$ for large modulations of the lattice. 
An investigation of this effect towards technological application will be part of
future work.

\end{document}